\documentclass[12pt]{article}
\usepackage{makeidx}
\usepackage{amssymb}
\usepackage{amsfonts}
\usepackage{amsmath}
\usepackage[T1]{fontenc}
\usepackage{graphicx}
\usepackage{setspace}
\usepackage{authblk}
\usepackage[left=2cm,top=2.5cm,right=2cm]{geometry}
\usepackage{units}
\usepackage{color}

\def\t{\textstyle}        
\def\one{1\hskip-.37em 1}                

\def\tH{{\widetilde{\cal H}}}

\def\H{{\cal H}}

\def\H{{\cal H}}

\def\wn{{\widetilde n}}

\def\ra{\rightarrow}

\def\tint{{\textstyle\int}}

\def\s{\hskip.08em}
\def\d{\partial}

\def\a{\alpha}
\def\b{\begin{eqnarray*}}  
\def\e{\end{eqnarray*}}    
\def\bn{\begin{eqnarray}}  
\def\en{\end{eqnarray}}   
\def\<{\langle}
\def\>{\rangle}

\def\no{\nonumber}

\def\{{\lbrace}
\def\}{\rbrace}

\begin{document}

\title{\textbf{Examples of Enhanced Quantization: \\ Bosons, Fermions, and Anyons}}

\author[1,2]{T. C. Adorno\thanks{tadorno@usp.br, tadorno@ufl.edu}}
\author[2,3]{J. R. Klauder\thanks{klauder@phys.ufl.edu}}
\affil[1]{\textit{Instituto de F\'{\i}sica, Universidade de S\~{a}o Paulo, Caixa Postal 66318, CEP 05508-090, S\~{a}o Paulo, S.P., Brazil;}}
\affil[2]{\textit{Department of Physics, University of Florida, 2001 Museum Road, Gainesville, FL 32611, USA;}}
\affil[3]{\textit{Department of Mathematics, University of Florida, 1400 Stadium Rd, Gainesville, FL 32611, USA.}}

\maketitle

\onehalfspacing

\begin{abstract}
Enhanced quantization offers a different classical/quantum connection than that of canonical quantization in which $\hbar>0$ throughout. This result arises when the only allowed Hilbert space vectors allowed in the quantum action functional are coherent states, which leads to the classical action functional augmented by additional terms of order $\hbar$. Canonical coherent states are defined by unitary transformations of a fixed, fiducial vector. While Gaussian vectors are commonly used as fiducial vectors, they cannot be used
for all systems. We focus on choosing fiducial vectors for several systems including bosons, fermions, and anyons.
\end{abstract}

\newpage

\section{Introduction}
Canonical quantization would seem to be, as its name suggests, a closed subject. However, some attempts at canonical quantization lead to unnatural results---most notably the ``Triviality of $\phi^4_n$ for $n\ge5$'', and possibly for $n=4$ as
well. Rather than accept such an outcome, a natural result may perhaps be obtained with an alternative quantization procedure, which, for the $\phi^4_n$ example, has been presented elsewhere \cite{cc}. The basis for an alternative quantization procedure \cite{EQ} is readily summarized here;
 for some early motivation, see \cite{Kla63I,Kla63II,Kla63III}. For the simple example of a single particle, the classical action functional is given by
  \bn A_C=\tint_0^T\s[\s p(t){\dot q}(t)-H_c(p(t),q(t)\s]\,dt\;,\en
  while the corresponding quantum action functional is given by
  \bn A_Q=\tint_0^T\s\<\psi(t)|\s[\s i\hbar(\d/\d t)-\H(P,Q)\s]\s|\psi(t)\>\,dt\;.\label{e2}\en
  A stationary variation of the paths $p(t)$ and $q(t)$ leads to Hamilton's equations of motion, while a stationary variation of normalized Hilbert state vectors $\<\psi(t)|$ (and $|\psi(t)\>$) leads to Schr\"odinger's equation of motion (and its adjoint). The connection between these two theories---known as ``canonical quantization''---comes  when we promote $p\ra P$ and $q\ra Q$ as Hermitian operators, and $\H(P,Q)$ is chosen as $H_c(P,Q)$ apart from possible ${\cal O}(\hbar)$ corrections. This prescription works best when $(p,q)$ are Cartesian coordinates \cite{dirac}---despite the fact that phase space carries no metric with which to determine Cartesian coordinates!

  There are many other proposed connections between classical and quantum expressions. We consider just two examples: (1) The Wigner \cite{Wigner} phase-space representation of a quantum state leads to a distribution function, but it generally suffers from not being a positive distribution; (2)
  The Husimi \cite{Husimi} phase-space representation provides a positive representation of an associated phase-space distribution function, but its partners in forming expectations are often singular distributions \cite{Sudar}. Further examples are presented in most textbooks dealing with quantum theory, e.g., \cite{BerShu91}.

Let us examine an alternative quantization procedure which has been called ``Enhanced Quantization'' \cite{EQ}.
The derivation of Schr\"odinger's  equation of motion assumes that sufficiently many vectors $\<\psi(t)|$ can be varied in (\ref{e2}), but suppose that is not possible. A macroscopic observer can only change a few features of a quantum system, such as its position or velocity, which thanks to Galilean covariance, can be realized by moving the observer rather than disturbing the microscopic system. Thus, choosing a suitable, normalized fiducial vector $|\eta\>$, and with two self-adjoint operators $P$ and $Q$, $[Q,P]=i\hbar$, and initially choosing $|\eta\>$  as $|0\>$, an harmonic oscillator ground state determined by $(Q+iP)\s|0\>=0$, we generate two, independent unitary transformations of $|0\>$ leading to
  \bn  |p,q\>\equiv e^{\t -iqP/\hbar}\,e^{\t ip\s Q/\hbar}\s |0\>\;. \label{coherstates}\en
We choose this set of states for all $(p,q)\in{\mathbb R}^2$ as states that can be varied by a macroscopic observer. This set is also well known as a set of canonical coherent states with $|0\>$ serving as its fiducial vector \cite{jkbs}. Using these states, the restricted quantum action functional becomes
   \bn A_{Q(R)}\hskip-1.3em&&=\tint_0^T\s\<p(t),q(t)|\s[\s i\hbar(\d/\d t)-\H(P,Q)\s]\s|p(t),q(t)\>\,dt\no\\
      &&=\tint_0^T\s[\s p(t){\dot q}(t)-H(p(t),q(t))\s]\,dt\;.\en
  Thus a stationary variation of the restricted action leads to Hamilton's equations with $H(p,q)$ serving as the enhanced classical Hamiltonian since $\hbar>0$ still. Specifically,
   \bn H(p,q)\hskip-1.3em&&\equiv \<p,q|\s\H(P,Q)\s|p,q\>\no\\
      &&=\<0|\s \H(P+p,Q+q)\s|0\>\no\\
    &&=\H(p,q)+{\cal O}(\hbar; p,q; |0\>)\;.\label{kla5} \en

It is important to note that this classical/quantum relation is exactly what is meant by the relation between the classical and quantum Hamiltonians when $(p,q)$ are the favored Cartesian coordinates. Although the phase space cannot determine Cartesian coordinates, the Hilbert space provides a metric. Since the overall phase of a Hilbert space vector has no physical significance in quantum theory, we use a suitably scaled version of the distance determined by $D_{ray}(|\psi\>;|\phi\>)^2\equiv(2\hbar)\min_\a \||\psi\>-e^{i\a}\s|\phi\>\|^2$, which, when applied to two infinitesimally close coherent states, leads to
\bn d\sigma(p,q)^2\equiv(2\hbar)\,[\s\|\s\s d|p,q\>\s\|^2-|\<p,q|\s\s d|p,q\>|^2\s]= dp^2+dq^2\;. \en
Thus, it is the Hilbert space that determines that the chosen coordinates are Cartesian, and, if one so desired, that metric could be added to the phase space as well. In that sense the alternative classical/quantum connection used in enhanced quantization has nevertheless yielded the very same result as canonical quantization since the Cartesian coordinates determined by the Hilbert space metric are linked with the very same choice of the Hamiltonian operator as related to the classical Hamiltonian as is chosen by canonical quantization.

There are additional features of enhanced quantization that are covered elsewhere, e.g., \cite{EQ, Moscow, Ben}.
They also include affine quantization, which although not discussed in this article, could also be related to our further discussions. In fact, the main role of this article is to feature the choice of the fiducial vector $|\eta\>$. So far we have chosen $|0\>$ as the fiducial vector, as this is a common choice whenever discussing coherent states. On the other hand, the choice of $|0\>$ is not always optimal or even possible, and in the rest of this paper we discuss the features and requirements of various fiducial vectors $|\eta\>$.

\section{Choosing the Fiducial Vector}
\subsection{General properties}
Coherent states are often defined with the aid of a group, and in so doing the action of the group acting on a fixed, fiducial vector defines the coherent states. As an example consider the set of canonical coherent states  given by
  \bn |p,q;\eta\>\equiv e^{\t-iqP/\hbar}\,e^{\t ipQ/\hbar}\,|\eta\>\;. \en
  Normally, the choice of the normalized fiducial vector $|\eta\>$ is left implicit, but on this occasion, since we are deciding on how to choose a ``good''---or the ``best''---fiducial vector, we include it explicitly in the previous equation. The transformation of the fiducial vector to make coherent states involves unitary transformations
  by the given expressions, which means that both P and Q must be self-adjoint operators so they may generate unitary operators. The real variables $p$ and $q$ each generate one-parameter groups expressed in so-called canonical group
  coordinates of the second kind \cite{Cohen}. Since such unitary operators are strongly continuous in their parameters, e.g., $\|\s[\s e^{ipQ/\hbar}-\one\s]\s|\psi\>\s\|\ra0$ as $p\ra0$ for all $|\psi\>\in{\mathfrak{ H}}$, it follows that the coherent states are strongly continuous in their parameters for any $|\eta\>$. This property ensures the continuity of the coherent state representation for any abstract vector $|\psi\>$ given by $\psi(p,q;\eta)\equiv\<p,q;\eta\s|\psi\>$ for any $|\eta\>$.

  Besides continuity, the other basic feature of coherent states is a resolution of unity by an integral over the entire phase space of coherent state projection operators that involves an absolutely continuous measure with a suitable positive weighting, which, for the example under consideration, is given by
     \bn \one=\int |p,q;\eta\>\s\<p,q;\eta|\,d\mu(p,q)\,,\hskip3em
     d\mu(p,q)\equiv dp\s\s dq/(2\pi\hbar)\;, \en
     a relation that holds weakly (as well as strongly) for any choice of the normalized fiducial vector $|\eta\>$. [Remark: For other sets of putative coherent states, it sometimes happens that the resolution of unity fails; in this case we deal with so-called ``weak coherent states'' \cite{jkbs}. When that happens, it is useful to let the inner product of weak coherent states serve as a reproducing kernel and to generate a reproducing kernel Hilbert space \cite{repro}.]

\subsection{First look at choosing the fiducial vector}
     In enhanced quantization, such as discussed in Sec.~1 of this paper, the restricted quantum action functional
     is given by
       \bn A_{Q(R)}\hskip-1.3em&&=\tint_0^T\<\,p,q;\eta\s|\s[\s i\hbar(\d/\d t)-\H(P,Q)\s]\s|p,q;\eta\>\,dt\no\\
        && =\tint_0^T[\s p(t){\dot q}(t)-H(p(t),q(t))\s]\,dt\;.\en
        and thus it is necessary that the the coherent states are in the domain of the Hamiltonian operator $\H$, which, for an unbounded Hamiltonian operator, will already induce a certain restriction on $|\eta\>$. Additionally, in giving physical meaning to the variables $p$ and $q$, it is useful to impose ``physical centering'', i,e,
        $\<P\>=\<Q\>=0$, wherein we have introduced the shorthand that $\<(\cdot)\>\equiv\<\eta|(\cdot)|\eta\>$.
        The virtue of physical centering becomes clear when we note that it leads to
          \bn \<p,q;\eta|\s P\s|p,q;\eta\>=p\;,\hskip3em \<p,q;\eta|\s Q\s|p,q;\eta\>=q\;, \en
          yielding a natural physical interpretation of the parameters $p$ and $q$.

          The enhanced classical Hamiltonian $H(p,q)$ (with $\hbar>0$) differs from the classical Hamiltonian
          $H_c(p,q)=\lim_{\hbar\ra0} H(p,q)$ by terms of order $\hbar$. As an example, let $\H(P,Q)=P^2+Q^2+Q^4$.
          in which case
            \bn H(p,q)=p^2+q^2+q^4+6q^2\<Q^2\> +\<\s[P^2+Q^2+Q^4]\>\;,\label{11K}\en
            assuming for simplicity that odd expectations vanish. Apart from a constant, there is the term $6q^2\<Q^2\>$ which will modify the usual equations of motion and their solution. If we choose
            $\<x|\eta\>\propto \exp(-\omega\s x^2/2\hbar)$, it follows that $\<Q^2\>\propto \hbar$, meaning that the correction is important only when $p$ and $q$ are ``quantum-sized'' themselves. That limitation makes good sense since then the enhanced classical description is effectively unchanged for macroscopic motion, only showing quantum ``uncertainty'', i.e., dependence on $\omega$, for quantum-sized motion. Alternatively, we could in principle choose
            $\<x|\eta\>\propto \exp(-a\s x^2/2)$, where, say, $a=10^{-137}\s m^{-2}$ and independent of $\hbar$, which means that the  additional term would modify the quadratic term by a potentially huge amount even for macroscopic motions. Such a choice is mathematically possible, just as studying a simple harmonic oscillator with displacements on a planetary scale or energies equivalent to the mass energy of the Earth are mathematically possible. However, they are unphysical applications of the mathematical description of
            a simple harmonic oscillator. In a similar story, although it is mathematically possible to choose fiducial vectors so that $\<Q^2\>$ leads to macroscopic modifications, such a fiducial vector could never be physically realized. Thus we conclude that it is logical to choose the fiducial vector supported largely on a ``quantum-sized'' region. However, that still leaves open many possibilities. Indeed, Troung \cite{TTT} has showed that choosing the ground state of a quartic Hamiltonian as the fiducial vector can recast Schr\"odinger's equation into a new form that offers novel analysis options.
\subsection{Second look at choosing the fiducial vector}
            It is popular to choose the fiducial vector to be a ``quantum-sized'' Gaussian; indeed, we have done so in Sec.~1. This choice often leads to fairly simple analytic expressions, and sometimes plausible arguments can be advanced that even help choose the variance parameter for such a vector \cite{gazeau}. However, it is important to understand that such a choice is {\it not} suitable in all cases. Let us reexamine the old discussion about {\it ``The rest of the universe''} \cite{fey}. A single system seldom exists in isolation; instead, it is surrounded by other systems. If we can imagine one specific system, then it is possible to imagine $N$ independent, identical systems, and even infinitely many such systems, i.e., $N=\infty$. A system may involve several degrees of freedom; however, for clarity our basic system has a single degree of freedom.

            For example, consider the Hamiltonian operator $\H_{(N)}\equiv\Sigma_{n=1}^N\,\H(P_n,Q_n)$. which represents $N$ independent, identical copies of the ``original'' Hamiltonian $\H(P_1,Q_1)$ involving independent operator pairs $(P_n,Q_n)$. The first system (for $P_1$ and $Q_1$)
             is chosen as the ``physical'' one, while the other sub-systems are ``spectator'' systems. The coherent
             states we choose are for the physical system only, that is only for the {\it first} operator pair; specifically
              \bn |p,q;\eta\>=e^{\t -iqP_1/\hbar}\,e^{\t ipQ_1/\hbar}\,|\eta\>\;,\hskip3em
                 |\eta\>\equiv \otimes_{n=1}^N\s|\eta_n\>\;.\en
                 To preserve the equivalence of all subsystems we choose each $|\eta_n\>$ to be identical to one another, i.e., the same single-system fiducial vector. In this case it follows that
                 \bn A_{Q(R)}=\tint_0^T[\s p(t){\dot q}(t)-H_{(N)}(p(t),q(t)\s]\,dt\;,\en
                 where, in the present case,
                   \bn H_{(N)}(p,q)=H(p,q)+\Sigma_{n=2}^N\,\<H(P_n,Q_n)\>\;, \label{e3}\en
                   which differs by a constant from the single system story. So long as $N<\infty$, that constant is finite and not important. But, as $N\ra\infty$, and we eventually deal with an infinite number of
                   spectator systems, it becomes important that that constant must be zero. At this point in the argument we restrict attention to quantum systems that have a non-negative spectrum and a unique ground state, which we choose as $|\eta\>$, with an energy eigenvalue adjusted to be zero. If that is the case, then the added constant in (\ref{e3}) vanishes for all $N$ including $N=\infty$. It may be argued that we could choose $|\eta\>$, say, as the first excited state for each subsystem and subtract that energy to obtain a zero. However, that would imply, for $N=\infty$, that there were infinitely many energy levels with $-\infty$ for their energy level, which is clearly an unphysical situation. The remedy for that situation is to insist that the fiducial vector be chosen as the unique ground state of each system
                   with an energy level adjusted to vanish. This choice applies to the physical system, and to all of the spectator systems as well. Thus we can imagine any one of the $N$ identical systems being the physical one and
                   the remaining $N-1$ systems as spectators.

                   Moreover, we can imagine there are many other spectator systems different from the physical one we have chosen.
                   For example, suppose there is another multiple-system type, with a Hamiltonian $\tH_{(\widetilde{N})}=\Sigma_{\wn=1}^{\widetilde{N}}\tH(P_\wn,Q_\wn)$, that is also present. This operator, too, is assumed to have a unique ground state with zero energy eigenvalue. Thus this new Hamiltonian could be present in the overall Hamiltonian but it would contribute nothing to the restricted quantum action functional because coherent states for it have not been ``turned on''. Indeed there could be many such systems, even infinitely many such new (sub)systems.
                    This argument can be carried to yet new families of spectator systems all of which are there, just ``resting'', or ``hibernating'', in their own ground state, and contributing nothing to the restricted quantum action functional. In this fashion we have found how to include ``the rest of the universe'' in such a way that it makes no contribution whatsoever, just as if we had ignored it altogether at the beginning of the story.

                   This desirable property requires that we choose the fiducial vector as the {\it unique ground state of the system under consideration adjusted to have zero energy}, which for this esoteric exercise of dealing with the surroundings proved extremely convenient if not absolutely necessary. This choice of fiducial vector also eliminates any nonsense regarding intrusion of the micro world into the macro world as we argued above. Of course, concerns about the intrusion of the surroundings are not always necessary, and thus it is acceptable to consider other fiducial vectors that are ``close''  to the ground state in some unspecified way, if one so desires.

                   Finally, we need to comment on other model systems that have Hamiltonians, which (i), near a lower bound, have a continuous spectrum, or (ii) instead have a spectrum that is unbounded below. These are interesting mathematical models, but it it is difficult to find any real physical systems that have such features.

                   Having shown how we can, if necessary, deal with the rest of the universe, we revert to simple systems without concerning ourselves with such big issues. Thus, in what follows, we allow ourselves to consider a variety of useful fiducial vectors, particularly those where Gaussian form are not appropriate to describe certain physical systems, as in the case of fermions and anyons.

\section{Bosons, Fermions, and Anyons}

In this section we confine ourselves to $2+1$ spacetime dimensions so that
anyons can be treated as well as bosons and fermions, although the results and
calculations for bosons and fermions are not limited to $2+1$ dimensions;
generalizations to arbitrary spacetime dimensions for them are straightfoward.

Charged particles orbiting around a magnetic flux tube and interacting with
it have fractional statistics in $2+1$ dimensions \cite{Wilczek}. Such
composites, known as anyons \cite{Wilczek}, are basically characterized by
their peculiar statistics: under a half rotation centered on a magnetic
flux, leading to a particle permutation, the state of the system changes by
a complex phase factor,

\begin{align}
& \psi (\mathbf{r}_{2},\mathbf{r}_{1})=e^{i\alpha }\psi (\mathbf{r}_{1},%
\mathbf{r}_{2})\,,\,\,(0\leq \alpha <2\pi )\,,  \label{any1} \\
& \mathbf{r}_{1}=(x_{1},y_{1})\,,\,\,\mathbf{r}_{2}=(x_{2},y_{2})\,,  \notag
\end{align}%
rather than the standard bosons ($\alpha =0$) or fermions ($\alpha =\pi $)
statistics. Here and throughout in this paper $\mathbf{r}_{\sigma }$
designates the two-dimensional cartesian coordinates of the first ($\sigma
=1 $) particle and the second ($\sigma =2$) particle, respectively.

As has been noticed originally by Wilczek \cite{Wilczek}, this consideration
has a very clear mathematical explanation based on the fact that in two
space dimensions the rotation group $SO(2)$ is isomorphic with the Abelian
unitary group $U(1)$ whose representations are labeled by real numbers. Due
to the latter fact and the spin-statistics connection, the wave function of
these particles may admit an arbitrary phase under rotations in $2+1$
spacetime. More precisely, circling a magnetic flux in the same direction, by two half turns, the state of the system may not necessarily need to return
to its original state, meaning that now the system can be described by a
multivalued wave function.

Anyons have attracted much attention due to their own richness, both in
theoretical treatment of fundamental concepts, as well as in their physical
implications. Particles with fractional statistics were considered in the $%
O(3)\,\sigma$ model due to the existence of solitons \cite{WilZee83}, it also
has been shown that anyons can be described as ordinary particles interacting
with a Chern-Simons field \cite{IenLec90,IenLec92}. A relativistic wave
equation for anyons was formulated in \cite{JacNai91} and more recently in
\cite{HorPlyVal10}. Regarding the physical implications, anyons have a
central role in the explanation of the quantum Hall effect \cite{Laughlin88}%
, in high-$T_{c}$ superconductivity \cite{FetHanLau89}, and more recently in
topological quantum computation \cite{Sarma05,Collins06,Stern08}. For a
complete review underlying the fundamental theoretical descriptions and
applications, we recommend the reviews \cite{Stern08,IenLec92,Wilczek90}.

All this interest motivates us to study the subject. Particularly, we are
interested in the enhanced classical theory of anyons in $2+1$ dimensions.
In this respect it should be noted that classical theories for
anyons have been considered before \cite{Pluyshchay90,Ghosh9495}. In these
papers the authors follow a canonical quantization procedure \cite%
{dirac,CanQuant} to derive the corresponding quantum theory for anyons.
Here, we adopt a different construction to quantize the theory of
particles with fractional statistics by the application of the coherent
state quantization and derive, for the first time, the corresponding enhanced
classical theory. More precisely we consider a rotationally invariant
Hamiltonian operator with a quartic interaction as a basic example.

Beneath the physical consideration behind anyons in $2+1$ dimensions, a
coherent state representation \cite{Kla63I,Kla63II} of the quantum theory
for anyons must exhibit the same property as (\ref{any1}). A direct
consequence of such a requirement is the statistical invariance of the
corresponding representation since now the fiducial vectors also obey,

\begin{equation}
\left\langle \mathbf{r}_{2},\mathbf{r}_{1}|\eta \right\rangle =\eta \left(
\mathbf{r}_{2},\mathbf{r}_{1}\right) =e^{i\alpha }\eta \left( \mathbf{r}_{1},%
\mathbf{r}_{2}\right) \,.  \label{any2}
\end{equation}

In choosing a coherent state representation we must, as a starting point,
provide a suitable fiducial vector consistent with the property (\ref{any2}%
). One possible way is to assume the fiducial vector is composed by two
parts $\eta \left( \mathbf{r}_{1},\mathbf{r}_{2}\right) =A\left( \mathbf{r}%
_{1},\mathbf{r}_{2}\right) S\left( \mathbf{r}_{1},\mathbf{r}_{2}\right) $
where $S\left( \mathbf{r}_{1},\mathbf{r}_{2}\right) $ is a symmetrical
function and $A\left( \mathbf{r}_{1},\mathbf{r}_{2}\right) $ has a nonsymmetrical form. In particular,%
\begin{equation*}
A\left( \mathbf{r}_{2},\mathbf{r}_{1}\right) =\left( -1\right) ^{\gamma
}A\left( \mathbf{r}_{1},\mathbf{r}_{2}\right) \,,
\end{equation*}%
and the condition (\ref{any2}) is recovered for $\gamma =\alpha /\pi $. This
construction leads to a proper description of nonrelativistic fermions for $%
\gamma =1$, and it is generalized for anyons assuming $0<\gamma <2\,\,
(\gamma\neq 1) $.

The wave functions for two or more anyon systems have been discussed before.
One of the first proposals was a multivalued function with a complex
nonsymmetrical part and a Gaussian-like symetrical part \cite%
{Halperin84,Wu84}. Generalizations of wave functions of this form have been
considered in several works (e.g. \cite%
{Chou91,MurLawBraBha91,IenLec92,Poly91}) and in particular represents a bound state
for two-anyons system in a $\left( 2+1\right) $-dim. interaction to an
external harmonic potential like $\left( 1/2\right) \Omega ^{2}\mathbf{r}%
_{i}^{2}$ \cite{Poly91}. Following these constructions, we study a family of
fiducial vectors, parametrized by $\lambda $, suitable to discuss bosons,
fermions, and anyons in the same framework,%
\begin{eqnarray}
&&\eta _{1}\left( \mathbf{r}_{1},\mathbf{r}_{2}\right) =N_{\gamma }\left(
z_{1}-z_{2}\right) ^{\gamma }e^{-\frac{\lambda }{2}\left( \mathbf{r}_{1}^{2}+%
\mathbf{r}_{2}^{2}\right) }\,,\ \ 0\leq \gamma <2\,,\ \ \gamma =\frac{\alpha
}{\pi }\,,  \label{any3} \\
&&z_{1}=x_{1}+iy_{1}\,,\ \ z_{2}=x_{2}+iy_{2}\,,\ \ \mathbf{r}_{\sigma
}^{2}=x_{\sigma }^{2}+y_{\sigma }^{2}\,,\ \ \lambda =\frac{\Omega }{\hbar }%
\,,\ \ \Omega =\mathrm{const.}\,,  \notag
\end{eqnarray}%
where $z_{1}$,$\ z_{2}$ are position of particles in the complex notation, $%
\Omega $ is an arbitrary real constant and $N_{\gamma }$ are the
corresponding normalization constants. It is clear that interchanging $%
z_{1}$ and $z_{2}$ is equivalent to changing the corresponding positions
of the particles, and, as a result, one obtains the desired phase $\eta
\left( \mathbf{r}_{2},\mathbf{r}_{1}\right) =\left( -1\right) ^{\gamma }\eta
\left( \mathbf{r}_{1},\mathbf{r}_{2}\right) $. We initially study
bosons and fermions in the next subsection. Exact solutions and results for
anyons are presented afterwards.

\subsection{General calculations for bosons and fermions}

In this subsection we examinate, in detail, the choice of fiducial vectors
(\ref{any3}) for the particular cases of bosons $\left( \gamma =0\right) $
and fermions $\left( \gamma =1\right) $. More precisely, we are interested to
obtain the enhanced classical description for these particles with bosons and
fermions under an influence of a quartic
interaction as (\ref{11K}). To do it one has to evaluate expectation values
of the self-adjoint operators $Q$, $P$, $Q^{2}$, $P^{2}$ and $Q^{4}$, and the
subsequent subsections are reserved to present that. Detailed derivations
are placed in the Appendix.

\subsubsection{Bosons $\left( \protect\gamma =0\right) $}

From (\ref{any3}) the fiducial vectors for bosons are pure Gaussian
functions,%
\begin{equation}
\eta _{1}\left( \mathbf{r}_{1},\mathbf{r}_{2}\right) =N_{0}e^{-\frac{\lambda
}{2}\left( \mathbf{r}_{1}^{2}+\mathbf{r}_{2}^{2}\right) }\,,\ \ N_{0}=\frac{%
\lambda }{\pi }\,.  \label{any4}
\end{equation}

One of the useful conditions underlying usual coherent state representations
lies in the fact that expectation values for the position and momentum
operators are zero. Due to parity properties one can see that $\left\langle
Q_{x_{i}}\right\rangle =0$ for the functions (\ref{any4}). Moreover it is
straightfoward to see that $\left\langle Q_{x_{i}}^{2n+1}\right\rangle =0$
for $n\in \mathbb{N}$. The expectation value of the momentum operator is
equivalently zero since the function (\ref{any4}) is symmetric in these
variables. Therefore,%
\begin{equation}
\left\langle Q_{x_{i}}\right\rangle =0=\left\langle P_{x_{i}}\right\rangle
\,.  \label{any6}
\end{equation}%
Intrinsic to the calculation of the enhanced Hamiltonian are the expectation
values of $Q_{x_{i}}^{2}$, $P_{x_{i}}^{2}$ and $Q_{x_{i}}^{4}$. For example,
the expectation value of $Q_{x_{1}}^{2}$ is,%
\begin{equation}
\left\langle Q_{x_{1}}^{2}\right\rangle =N_{0}^{2}\left( \frac{\pi }{\lambda
}\right) \int_{0}^{2\pi }d\vartheta \cos ^{2}\vartheta \int_{0}^{\infty
}dr_{1}r_{1}^{3}e^{-\lambda r_{1}^{2}}=\frac{1}{2\lambda }=\frac{\hbar }{%
2\Omega }\,,  \label{any7.1}
\end{equation}%
where $\vartheta $ is the polar angle. [Remark: Here and in what follows we adopt the convention that one integration symbol
over a bold letter means a double integration over two coordinates, for
example, $\int_{c}^{c^{\prime }}d\mathbf{r}f\left( x,y\right)\equiv\int_{c}^{c^{\prime }}dx\int_{c}^{c^{\prime }}dyf\left( x,y\right) $]. In addition the expectation value for
$P_{x_{1}}^{2}$ is
\begin{equation}
\left\langle P_{x_{1}}^{2}\right\rangle =-\hbar ^{2}\lambda
N_{0}^{2}\int_{-\infty }^{\infty }d\mathbf{r}_{1}\int_{-\infty }^{\infty }d%
\mathbf{r}_{2}\left( \lambda x_{1}^{2}-1\right) e^{-\lambda \left( \mathbf{r}%
_{1}^{2}+\mathbf{r}_{2}^{2}\right) }=\frac{\hbar ^{2}\lambda }{2}=\frac{%
\hbar \Omega }{2}\,,  \label{any7.2}
\end{equation}%
and the remaining expectation values have the same result.

In order to discuss the enhanced classical theory for bosons with a quartic
interaction one has to evaluate several expectation values like $%
\left\langle Q_{x_{\sigma }}^{4}\right\rangle $, $\langle Q_{x_{\sigma
}}^{2}Q_{x_{\sigma ^{\prime }}}^{2}\rangle $ and $\langle
Q_{x_{\sigma }}^{2}Q_{y_{\sigma ^{\prime }}}^{2}\rangle $. However,
due to the symmetry of the Gaussian fiducial vectors (\ref{any4}), only $%
\left\langle Q_{x_{1}}^{4}\right\rangle $ is independent,%
\begin{equation}
\left\langle Q_{x_{1}}^{4}\right\rangle =N_{0}^{2}\left( \frac{\pi }{\lambda
}\right) \int_{0}^{2\pi }d\vartheta \cos ^{4}\vartheta \int_{0}^{\infty
}dr_{1}r_{1}^{5}e^{-\lambda r_{1}^{2}}=\frac{3}{4\lambda ^{2}}=\frac{3\hbar
^{2}}{4\Omega ^{2}}\,.  \label{any7.3}
\end{equation}%
To show that consider, for example, $\left\langle
Q_{x_{1}}^{2}Q_{x_{2}}^{2}\right\rangle $ (which is equivalent to $%
\left\langle Q_{x_{1}}^{2}Q_{y_{1}}^{2}\right\rangle $ in the present case).
From its form it is straightfoward to conclude that%
\begin{equation}
\left\langle Q_{x_{1}}^{2}Q_{x_{2}}^{2}\right\rangle =N_{0}^{2}\left( \frac{%
\pi }{\lambda }\right) \int_{-\infty }^{\infty }dx_{1}x_{1}^{2}e^{-\lambda
x_{1}^{2}}\int_{-\infty }^{\infty }dx_{2}x_{2}^{2}e^{-\lambda x_{2}^{2}}=%
\frac{\left\langle Q_{x_{1}}^{4}\right\rangle }{3}\,.  \label{any7.4}
\end{equation}

\subsubsection{Fermions $\left( \protect\gamma =1\right)$}

For fermions the fiducial vectors (\ref{any3}) take the form%
\begin{equation}
\eta _{1}\left( \mathbf{r}_{1},\mathbf{r}_{2}\right) =N_{1}\left(
z_{1}-z_{2}\right) e^{-\frac{\lambda }{2}\left( \mathbf{r}_{1}^{2}+\mathbf{r}%
_{2}^{2}\right) }\,,  \label{fer1.1}
\end{equation}%
and the corresponding normalization constant is obtained as usual,%
\begin{equation}
\left\vert N_{1}\right\vert =\left( \int_{-\infty }^{\infty }d\mathbf{r}%
_{1}\int_{-\infty }^{\infty }d\mathbf{r}_{2}\left\vert \mathbf{r}_{1}-%
\mathbf{r}_{2}\right\vert ^{2}e^{-\lambda \left( \mathbf{r}_{1}^{2}+\mathbf{r%
}_{2}^{2}\right) }\right) ^{-\frac{1}{2}}=\sqrt{\frac{\lambda ^{3}}{2\pi ^{2}%
}}\,,  \label{fer2.1}
\end{equation}%
The expectation value of the coordinate operators, $Q_{x_{1}},...$ etc, are
zero for these states. For example, $\left\langle Q_{x_{1}}\right\rangle $
has the form%
\begin{equation}
\left\langle Q_{x_{1}}\right\rangle =N_{1}^{2}\int_{0}^{2\pi }d\vartheta
\cos \vartheta \int_{0}^{\infty }dr_{1}r_{1}^{2}e^{-\lambda
r_{1}^{2}}\int_{-\infty }^{\infty }d\mathbf{r}_{2}\left\vert \mathbf{r}_{1}-%
\mathbf{r}_{2}\right\vert ^{2}e^{-\lambda \mathbf{r}_{2}^{2}}=0\,,
\label{fer3}
\end{equation}%
In addition the expectation values for the momentum operator $\left\langle
P_{x_{1}}\right\rangle $ reads,%
\begin{equation}
\left\langle P_{x_{1}}\right\rangle =-i\hbar N_{1}^{2}\int_{-\infty
}^{\infty }d\mathbf{r}_{1}\int_{-\infty }^{\infty }d\mathbf{r}_{2}\left\{
\left( z_{1}^{\ast }-z_{2}^{\ast }\right) -\lambda x_{1}\left\vert \mathbf{r}%
_{1}-\mathbf{r}_{2}\right\vert ^{2}\right\} e^{-\lambda \left( \mathbf{r}%
_{1}^{2}+\mathbf{r}_{2}^{2}\right) }=0\,,  \label{fer4}
\end{equation}%
and the remaining linear expectation values are also zero. The expectation value
for $Q_{x_{1}}^{2}$ are%
\begin{equation}
\left\langle Q_{x_{1}}^{2}\right\rangle =N_{1}^{2}\int_{-\infty }^{\infty }d%
\mathbf{r}_{1}\int_{-\infty }^{\infty }d\mathbf{r}_{2}\left( \mathbf{r}%
_{1}^{2}+\mathbf{r}_{2}^{2}\right) x_{1}^{2}e^{-\lambda \left( \mathbf{r}%
_{1}^{2}+\mathbf{r}_{2}^{2}\right) }=\frac{3}{4\lambda }=\frac{3\hbar }{%
4\Omega }\,,  \label{fer5.0}
\end{equation}%
and for $P_{x_{1}}^{2}$ are%
\begin{eqnarray}
\left\langle P_{x_{1}}^{2}\right\rangle &=&-\hbar ^{2}\lambda
N_{1}^{2}\int_{-\infty }^{\infty }d\mathbf{r}_{1}\int_{-\infty }^{\infty }d%
\mathbf{r}_{2}\left\{ \left\vert \mathbf{r}_{1}-\mathbf{r}_{2}\right\vert
^{2}\left( \lambda x_{1}^{2}-1\right) -x_{1}^{2}\right\} e^{-\lambda \left(
\mathbf{r}_{1}^{2}+\mathbf{r}_{2}^{2}\right) }  \label{fer6.0} \\
&=&\frac{3\hbar ^{2}\lambda }{4}=\frac{3\hbar \Omega }{4}\,,  \notag
\end{eqnarray}

In order to discuss rotationally invariant quartic interactions we have to
evaluate, for example, expectation values as $\left\langle Q_{x_{\sigma
}}^{4}\right\rangle $, $\langle Q_{x_{\sigma }}^{2}Q_{x_{\sigma
^{\prime }}}^{2}\rangle $ and $\langle Q_{x_{\sigma
}}^{2}Q_{y_{\sigma ^{\prime }}}^{2}\rangle $. The first one has the
form%
\begin{equation}
\left\langle Q_{x_{1}}^{4}\right\rangle =N_{1}^{2}\int_{-\infty }^{\infty }d%
\mathbf{r}_{1}\int_{-\infty }^{\infty }d\mathbf{r}_{2}\left\vert \mathbf{r}%
_{1}-\mathbf{r}_{2}\right\vert ^{2}x_{1}^{4}e^{-\lambda \left( \mathbf{r}%
_{1}^{2}+\mathbf{r}_{2}^{2}\right) }=\frac{3}{2\lambda ^{2}}=\frac{3\hbar
^{2}}{2\Omega ^{2}}\,,  \label{fer7}
\end{equation}%
and the same result can be obtained for $\left\langle
Q_{y_{1}}^{4}\right\rangle $ ... etc. Next it can easely seen that
expectation values like $\langle Q_{x_{\sigma }}^{2}Q_{x_{\sigma
^{\prime }}}^{2}\rangle $ and $\langle Q_{x_{\sigma
}}^{2}Q_{y_{\sigma ^{\prime }}}^{2}\rangle $ enjoy the same property
as (\ref{any7.4}). At last, there exists one more interesting expectation
value $\left\langle Q_{x_{1}}Q_{x_{2}}\right\rangle $ which vanishes for
bosons but is nonzero for fermions,%
\begin{equation}
\left\langle Q_{x_{1}}Q_{x_{2}}\right\rangle =-2N_{1}^{2}\int_{-\infty
}^{\infty }d\mathbf{r}_{1}\int_{-\infty }^{\infty }d\mathbf{r}%
_{2}x_{1}x_{2}\left( \mathbf{r}_{1}\cdot \mathbf{r}_{2}\right) e^{-\lambda
\left( \mathbf{r}_{1}^{2}+\mathbf{r}_{2}^{2}\right) }=-\frac{1}{4\lambda }=-%
\frac{\hbar }{4\Omega }\,.  \label{fer8}
\end{equation}%
which again, due to the symmetry of the problem, this result coincides with $%
\left\langle Q_{y_{1}}Q_{y_{2}}\right\rangle $.

\subsection{General calculations for anyons}

In this section we generalize the previous results for arbitrary $\gamma $
within the range $\left[ 0,2\right) $, with the $\left( 0,1\right) \cup
(1,2) $ used for anyons. More precisely, the choice of fiducial vector (\ref%
{any3}) admits an exact solution for arbitrary $\gamma $ and the results here
are an extension to those particular cases with $\gamma =0$ and $1$. The
first step is to evaluate the norm of the corresponding fiducial vectors,%
\begin{equation}
\left\Vert \left\vert \eta \right\rangle \right\Vert ^{2}=\left\vert
N_{\gamma }\right\vert ^{2}\int_{-\infty }^{\infty }d\mathbf{r}%
_{1}\int_{-\infty }^{\infty }d\mathbf{r}_{2}\left\vert \mathbf{r}_{1}-%
\mathbf{r}_{2}\right\vert ^{2\gamma }e^{-\lambda \left( \mathbf{r}_{1}^{2}+%
\mathbf{r}_{2}^{2}\right) }\,.  \label{any8}
\end{equation}%
where again the normalization constant is obtained as usual%
\begin{equation}
\left\vert N_{\gamma }\right\vert =\left( \int_{-\infty }^{\infty }d\mathbf{r%
}_{1}e^{-\lambda \mathbf{r}_{1}^{2}}\int_{-\infty }^{\infty }d\mathbf{r}%
_{2}\left\vert \mathbf{r}_{1}-\mathbf{r}_{2}\right\vert ^{2\gamma
}e^{-\lambda \mathbf{r}_{2}^{2}}\right) ^{-\frac{1}{2}}\,,  \label{any8.1}
\end{equation}%
This rotationally-invariant integral can simplified by choosing a
convenient coordinate system. For example, in the case where $\mathbf{r}_{2}$
is aligned along the axis $y_{1}$ (\ref{any8.1}) has the form,%
\begin{eqnarray}
&&\left\vert N_{\gamma }\right\vert ^{2}=\left( \int_{-\infty }^{\infty }d%
\mathbf{r}_{2}e^{-\lambda \mathbf{r}_{2}^{2}}\rho _{\gamma }^{\left(
-\right) }\left( r_{2}\right) \right) ^{-1}\,,\ \ \rho _{\gamma }^{\left(
-\right) }\left( r_{2}\right) \equiv \int_{0}^{\infty
}dr_{1}r_{1}e^{-\lambda r_{1}^{2}}\mathcal{I}_{\vartheta }^{\left( -\right)
}\left( r_{1},r_{2};\gamma \right) \,,  \notag \\
&&\mathcal{I}_{\vartheta }^{\left( -\right) }\left( r_{1},r_{2};\gamma
\right) \equiv \int_{0}^{2\pi }d\vartheta \left(
r_{1}^{2}+r_{2}^{2}-2r_{1}r_{2}\sin \vartheta \right) ^{\gamma }\,,
\label{any9}
\end{eqnarray}%
where $\vartheta $ is the angle between the axis $x_{1}$ and $\mathbf{r}_{1}$ such that $\mathbf{r}_{1}\cdot \mathbf{r}_{2}=r_{1}r_{2}\cos \left( \pi
/2-\vartheta \right) =r_{1}r_{2}\sin \vartheta $. Detailed solution of $\rho _{\gamma }^{\left( -\right)
}\left( r_{2}\right) $ and subsequent calculations can be found at the
Appendix (\ref{ap1})-(\ref{ap5.1}). Using those results the norm (\ref{any8}%
) has the final form%
\begin{equation}
\left\Vert \left\vert \eta \right\rangle \right\Vert ^{2}=\left\vert
N_{\gamma }\right\vert ^{2}\frac{\pi ^{2}\sqrt{\pi }}{2\lambda ^{\gamma +2}}%
\frac{\Gamma \left( 2+\gamma \right) }{\Gamma \left( \frac{3}{2}\right) }%
\left. _{2}F_{1}\left( \frac{1-\gamma }{2},-\frac{\gamma }{2};\frac{3}{2}%
;1\right) \right. \,,  \label{any14}
\end{equation}%
from which follows that%
\begin{equation}
\left\vert N_{\gamma }\right\vert =\left[ \frac{\pi ^{2}\sqrt{\pi }}{%
2\lambda ^{\gamma +2}}\frac{\Gamma \left( 2+\gamma \right) }{\Gamma \left(
\frac{3}{2}\right) }\left. _{2}F_{1}\left( \frac{1-\gamma }{2},-\frac{\gamma
}{2};\frac{3}{2};1\right) \right. \right] ^{-1/2}\,,  \label{any15}
\end{equation}
where $\left. _{2}F_{1}\left(\alpha ,\beta ;\delta ;z\right) \right. $ is the hypergeometric function \cite{Gradshtein}.
The expectation value $Q_{x_{1}}^{2}$ is conveniently evaluated in the same reference system, whose form is%
\begin{equation}
\left\langle Q_{x_{1}}^{2}\right\rangle =\left\vert N_{\gamma }\right\vert
^{2}\int_{-\infty }^{\infty }d\mathbf{r}_{1}e^{-\lambda \mathbf{r}%
_{1}^{2}}x_{1}^{2}\rho _{\gamma }^{\left( -\right) }\left( r_{1}\right) \,,
\label{any15.1}
\end{equation}%
where $\rho _{\gamma }^{\left( -\right) }\left( r_{1}\right) $ is defined in
(\ref{ap1}). Using its solution (\ref{ap4}) this expectation value can be
written as\footnote{%
For the explicit definition of $\Lambda _{s}^{\left( -\right) }$ and $%
R_{1}^{\left( -\right) }\left( \gamma ,s\right) $ see the Appendix (\ref{ap2}%
), (\ref{ap5.1}).},%
\begin{equation}
\left\langle Q_{x_{1}}^{2}\right\rangle =\frac{\pi ^{2}}{2\lambda ^{2}}%
\left\vert N_{\gamma }\right\vert ^{2}\sum_{s=0}^{\infty }\Lambda
_{s}^{\left( -\right) }R_{1}^{\left( -\right) }\left( \gamma ,s\right)
=\left( \frac{\pi ^{2}\sqrt{\pi }}{8\lambda ^{3+\gamma }}\right) \left\vert
N_{\gamma }\right\vert ^{2}\frac{\Gamma \left( 3+\gamma \right) }{\Gamma
\left( \frac{3}{2}\right) }\left. _{2}F_{1}\left( \frac{1-\gamma }{2},-\frac{%
\gamma }{2};\frac{3}{2};1\right) \right. \,,  \label{any16}
\end{equation}%
and substituting (\ref{any15}) we get a remarkably simple result:%
\begin{equation}
\left\langle Q_{x_{1}}^{2}\right\rangle =\frac{1}{4\lambda }\frac{\Gamma
\left( 3+\gamma \right) }{\Gamma \left( 2+\gamma \right) }=\frac{\left(
\gamma +2\right) }{4\lambda }=\frac{\hbar \left( \gamma +2\right) }{4\Omega }%
\,.  \label{any17}
\end{equation}

By the same symmetry arguments, which has been discussed in the section on bosons and fermions, one can see that $\left\langle Q_{x_{1}}\right\rangle =0$
(again, the integral with respect to $x_{1}$ has an odd integrand within a symmetric interval). This is also true for the remaining coordinates
expectation values. By virtue of this fact it follows that the expectation
values of the momentum operators are also zero, for example,%
\begin{eqnarray}
\left\langle P_{x_{1}}\right\rangle &=&\left\vert N_{\gamma }\right\vert
^{2}\int_{-\infty }^{\infty }d\mathbf{r}_{1}\int_{-\infty }^{\infty }d%
\mathbf{r}_{2}\left\{ \left( z_{1}^{\ast }-z_{2}^{\ast }\right) \left[
\gamma -\lambda x_{1}\left( z_{1}-z_{2}\right) \right] \right.  \notag \\
&\times &\left. \left( \mathbf{r}_{1}^{2}+\mathbf{r}_{2}^{2}-2\mathbf{r}%
_{1}\cdot \mathbf{r}_{2}\right) ^{\gamma -1}e^{-\lambda \left( \mathbf{r}%
_{2}^{2}+\mathbf{r}_{1}^{2}\right) }\right\} =0\,,  \label{any18}
\end{eqnarray}%
since it only has odd contributions.

The full expression of $\left\langle P_{x_{1}}^{2}\right\rangle $ is,%
\begin{eqnarray*}
\left\langle P_{x_{1}}^{2}\right\rangle &=&-\hbar ^{2}\left\vert N_{\gamma
}\right\vert ^{2}\int_{-\infty }^{\infty }d\mathbf{r}_{1}\int_{-\infty
}^{\infty }d\mathbf{r}_{2}e^{-\lambda \left( \mathbf{r}_{2}^{2}+\mathbf{r}%
_{1}^{2}\right) } \{\left\vert \mathbf{r}_{1}-\mathbf{r}_{2}\right\vert
^{2\left( \gamma -2\right) } \\
&\times &\left[ \gamma \left( \gamma -1\right) \left( z_{1}^{\ast
}-z_{2}^{\ast }\right) ^{2}-2\lambda \gamma x_{1}\left( z_{1}^{\ast
}-z_{2}^{\ast }\right) \left\vert \mathbf{r}_{1}-\mathbf{r}_{2}\right\vert
^{2}\right. \\
&+&\left. \left. \lambda \left( \lambda x_{1}^{2}-1\right) \left\vert
\mathbf{r}_{1}-\mathbf{r}_{2}\right\vert ^{4}\right] \right\} \,.
\end{eqnarray*}%
Due to the appearance of odd coefficients some terms in this integral
vanishe. Using previous results (\ref{any14}), (\ref{any16}) and the
definitions (\ref{ap1}), (\ref{ap3}), (\ref{ap5.1}), this expectation value
can be written as (replacing $\lambda $ by $\Omega /\hbar $),%
\begin{eqnarray}
\left\langle P_{x_{1}}^{2}\right\rangle &=&-\hbar ^{2}\left\vert N_{\gamma
}\right\vert ^{2}\sum_{s=0}^{\infty }\Lambda _{s}^{\left( -\right)
}R_{1}^{\left( -\right) }\left( \gamma -1,s\right) +\hbar ^{2}\lambda \left(
1-\lambda \left\langle Q_{x_{1}}^{2}\right\rangle \right)  \notag \\
&=&\frac{\Omega \hbar }{2}\left[ 1+\gamma \left( \frac{\left.
_{2}F_{1}\left( \frac{2-\gamma }{2},\frac{1-\gamma }{2};\frac{3}{2};1\right)
\right. }{\left. _{2}F_{1}\left( \frac{1-\gamma }{2},-\frac{\gamma }{2};%
\frac{3}{2};1\right) \right. }\right) -\frac{\gamma }{2}\right] \,.
\label{any21}
\end{eqnarray}%
It should be noted that both results (\ref{any17}) and (\ref{any21})
coincide with the particular considerations (\ref{any7.1}), (\ref{any7.2})
and (\ref{fer5.0}), (\ref{fer6.0}), as expected.

Among all possible rotationally-invariant potentials within the problem under
consideration are quartic interactions. There is a particular interest
in rotational invariant models, which justifies our study.
We consider the rotationally invariant potential $V$ given by,%
\begin{equation}
V\equiv \left( \mathbf{Q}_{1}^{2}+\mathbf{Q}_{2}^{2}\right) ^{2}\,,\ \
\mathbf{Q}_{1}^{2}\equiv Q_{x_{1}}^{2}+Q_{y_{1}}^{2}\,,\ \ \mathbf{Q}%
_{2}^{2}\equiv Q_{x_{2}}^{2}+Q_{y_{2}}^{2}\,,  \label{any22.1}
\end{equation}%
whose expectation value can separated into a homogenous part $V_{\mathrm{H}}$
and in a nonhomogeneous part $V_{\mathrm{N}}$,%
\begin{equation}
\left\langle V\right\rangle =\left\langle V_{\mathrm{H}}\right\rangle
+\left\langle V_{\mathrm{N}}\right\rangle \,,\ \ \left\langle V_{\mathrm{H}%
}\right\rangle \equiv\left\langle \mathbf{Q}_{1}^{4}+\mathbf{Q}%
_{2}^{4}\right\rangle \,,\ \ \left\langle V_{\mathrm{N}}\right\rangle
\equiv 2\left\langle \mathbf{Q}_{1}^{2}\mathbf{Q}_{2}^{2}\right\rangle \,.
\label{any22.2}
\end{equation}%
From its form it can be seen that the corresponding expectation value
contains terms like $\left\langle Q_{x_{1}}^{4}\right\rangle $ (the
same for all other coordinates operators), as well as $\left\langle
Q_{x_{1}}^{2}Q_{y_{1}}^{2}\right\rangle $, and correlation functions such as $%
\left\langle Q_{x_{1}}^{2}Q_{x_{2}}^{2}\right\rangle $. For example, the
expectation value of $Q_{x_{1}}^{4}$ is%
\begin{equation}
\left\langle Q_{x_{1}}^{4}\right\rangle =\left\vert N_{\gamma }\right\vert
^{2}\int_{-\infty }^{\infty }d\mathbf{r}_{1}e^{-\lambda \mathbf{r}%
_{1}^{2}}x_{1}^{4}\rho _{\gamma }^{\left( -\right) }\left( r_{1}\right) =%
\frac{3\pi }{8\lambda ^{3}}\left\vert N_{\gamma }\right\vert
^{2}\sum_{s=0}^{\infty }\Lambda _{s}^{\left( -\right) }R_{2}^{\left(
-\right) }\left( \gamma ,s\right) \,,  \label{any23}
\end{equation}%
which, after a few simplifications, can be split into two parts%
\begin{eqnarray*}
&&\left\langle Q_{x_{1}}^{4}\right\rangle =\left\vert N_{\gamma }^{\left(
1\right) }\right\vert ^{2}\left( \frac{3\pi ^{2}\sqrt{\pi }}{64\lambda
^{\gamma +4}}\right) \frac{\Gamma \left( \gamma +4\right) }{\Gamma \left(
5/2\right) }\left( \mathcal{F}_{1}+\mathcal{F}_{2}\right) \,, \\
&&\mathcal{F}_{1}=\sum_{s=1}^{\infty }\frac{1}{\left( s-1\right) !}\left(
\frac{1-\gamma }{2}\right) _{s}\left( -\frac{\gamma }{2}\right) _{s}\left(
\frac{5}{2}\right) _{s}^{-1}\,,\ \ \mathcal{F}_{2}=2\left. _{2}F_{1}\left(
\frac{1-\gamma }{2},-\frac{\gamma }{2};\frac{5}{2};1\right) \right. \,.
\end{eqnarray*}%
Using the standard relation $\left( a\right) _{s+1}=a\left( a+1\right) _{s}$ the
first series $\mathcal{F}_{1}$ can be simplified and be identified as a new
hypergeometric function,%
\begin{equation*}
\mathcal{F}_{1}=\frac{\gamma \left( \gamma -1\right) }{10}\left.
_{2}F_{1}\left( \frac{3-\gamma }{2},\frac{2-\gamma }{2};\frac{7}{2};1\right)
\right. \,,
\end{equation*}%
and consequently the expectation value of $\left\langle
Q_{x_{1}}^{4}\right\rangle $ has its final form (replacing $\lambda $ by $%
\Omega /\hbar $)%
\begin{eqnarray}
\left\langle Q_{x_{1}}^{4}\right\rangle &=&\frac{\hbar ^{2}\left( \gamma
+3\right) \left( \gamma +2\right) }{16\Omega ^{2}\left. _{2}F_{1}\left(
\frac{1-\gamma }{2},-\frac{\gamma }{2};\frac{3}{2};1\right) \right. }\left[
2\left. _{2}F_{1}\left( \frac{1-\gamma }{2},-\frac{\gamma }{2};\frac{5}{2}%
;1\right) \right. \right.  \notag \\
&+&\left. \frac{\gamma \left( \gamma -1\right) }{10}\left. _{2}F_{1}\left(
\frac{3-\gamma }{2},\frac{2-\gamma }{2};\frac{7}{2};1\right) \right. \right]
\,.  \label{any24}
\end{eqnarray}%
Although (\ref{any24}) has an apparently complicated form, one can see that
it reduces to (\ref{any7.3}) and (\ref{fer5.0}) when $\gamma =0$, $\gamma =1$%
, respectively. Since $\left. _{2}F_{1}\left( a,b;c;1\right) \right. =1$ for
$a=0$ or $b=0$, one can readily check that%
\begin{equation*}
\left. \left\langle Q_{x_{1}}^{4}\right\rangle \right\vert _{\gamma =0}=%
\frac{12\hbar ^{2}}{16\Omega ^{2}}=\frac{3\hbar ^{2}}{4\Omega ^{2}}\,,\ \
\left. \left\langle Q_{x_{1}}^{4}\right\rangle \right\vert _{\gamma =1}=%
\frac{24\hbar ^{2}}{16\Omega ^{2}}=\frac{3\hbar ^{2}}{2\Omega ^{2}}\,,
\end{equation*}%
as expected.

The second kind of interaction from (\ref{any22.1}) has the form $%
\left\langle Q_{x_{1}}^{2}Q_{y_{1}}^{2}\right\rangle $. It can be seen that
for this interaction%
\begin{equation}
\left\langle Q_{x_{1}}^{2}Q_{y_{1}}^{2}\right\rangle =\left\vert N_{\gamma
}\right\vert ^{2}\int_{-\infty }^{\infty }d\mathbf{r}_{1}e^{-\lambda \mathbf{%
r}_{1}^{2}}x_{1}^{2}y_{1}^{2}\rho _{\gamma }^{\left( -\right) }\left(
r_{1}\right) \,,  \label{any25.1}
\end{equation}%
differs from (\ref{any23}) only in the first integral which, after changing
to polar coordinates, one obtains the simple relation%
\begin{equation}
\left\langle Q_{x_{1}}^{2}Q_{y_{1}}^{2}\right\rangle =\frac{\left\langle
Q_{x_{1}}^{4}\right\rangle }{3}\,.  \label{any25.2}
\end{equation}%
It is worth noting that this result obeys the same standard property of pure
Gaussian states, i. e., (\ref{any25.2}) is true for Gaussian states and here
we see that this same property holds true. The reason behind it is due to
the fact that the \textquotedblleft non-Gaussian\textquotedblright\ part of
these expectation values is a rotational invariant function which, as a
matter of fact, does not spoil such a property. More specifically, the
property%
\begin{equation*}
\int_{-\infty }^{\infty }dx\int_{-\infty }^{\infty }dy\,e^{-\left(
x^{2}+y^{2}\right) }x^{2}y^{2}f\left( r\right) =\frac{1}{3}\int_{-\infty
}^{\infty }dx\int_{-\infty }^{\infty }dy\,e^{-\left( x^{2}+y^{2}\right)
}x^{4}f\left( r\right) \,,
\end{equation*}%
is true for any rotational invarian function $f\left( r\right) $.

At last, the remaining interactions are mixed and nonhomogeneous like $%
\left\langle Q_{x_{1}}^{2}Q_{x_{2}}^{2}\right\rangle $. These are slightly
different from the previous cases (\ref{any23}), (\ref{any25.1}) as one can
see from its form,%
\begin{eqnarray}
&&\left\langle Q_{x_{1}}^{2}Q_{x_{2}}^{2}\right\rangle =\left\vert N_{\gamma
}\right\vert ^{2}\int_{-\infty }^{\infty }d\mathbf{r}_{1}x_{1}^{2}e^{-%
\lambda \mathbf{r}_{1}^{2}}\rho _{\gamma }^{\left( +\right) }\left(
r_{1}\right) \,,\ \ \rho _{\gamma }^{\left( +\right) }\left( u_{1}\right)
=\int_{0}^{\infty }dr_{2}r_{2}^{3}e^{-\lambda r_{2}^{2}}\mathcal{I}%
_{\vartheta }^{\left( +\right) }\left( r_{1},r_{2};\gamma \right) \,,  \notag
\\
&&\mathcal{I}_{\vartheta }^{\left( +\right) }\left( r_{1},r_{2};\gamma
\right) =\int_{0}^{2\pi }d\vartheta \cos ^{2}\vartheta \left(
r_{1}^{2}+r_{2}^{2}-2r_{1}r_{2}\sin \vartheta \right) ^{\gamma }\,,
\label{any26.1}
\end{eqnarray}%
and the solution of $\mathcal{I}_{\vartheta }^{\left( +\right) }\left(
r_{1},r_{2};\gamma \right) $ can be found at the appendix (\ref{ap2}). Using
the series representation of the hypergeometric function as (\ref{ap2.1})
the integral $\rho _{\gamma }^{\left( +\right) }\left( u_{1}\right) $ can be
written as (\ref{ap3})%
\begin{eqnarray}
\rho _{\gamma }^{\left( +\right) }\left( u_{1}\right) &=&\sum_{s=0}^{\infty
}\Lambda _{s}^{\left( +\right) }\varrho _{\gamma ,s}^{\left( +\right)
}\left( u_{1}\right) u_{1}^{s}\,,\ \ \Lambda _{s}^{\left( +\right) }=\frac{%
\pi }{\lambda ^{\gamma +2}}\frac{4^{s}}{s!}\left( \frac{1-\gamma }{2}\right)
_{s}\left( -\frac{\gamma }{2}\right) _{s}\left( 2\right) _{s}^{-1}  \notag \\
\varrho _{\gamma ,s}^{\left( +\right) }\left( u_{1}\right) &=&\Gamma \left(
s+2\right) u_{1}^{\frac{\gamma -s+1}{2}}e^{\frac{u_{1}}{2}}W_{\left( \gamma
-3s-1\right) /2,\left( s-\gamma -2\right) /2}\left( u_{1}\right) \,,
\label{any26.2}
\end{eqnarray}%
where the formula (\ref{ap4}) has been used above. After some minor
algebraic manipulations and using (\ref{ap5.1}) we get the final result
(replacing $\lambda $ by $\Omega /\hbar $),%
\begin{equation}
\left\langle Q_{x_{1}}^{2}Q_{x_{2}}^{2}\right\rangle =\left( \frac{\hbar
^{2}\left( \gamma +3\right) \left( \gamma +2\right) }{24\Omega ^{2}}\right)\frac{
 \left. _{2}F_{1}\left( \frac{1-\gamma }{2},-\frac{\gamma }{2};\frac{5%
}{2};1\right) \right.}{ \left. _{2}F_{1}\left( \frac{1-\gamma }{2},-%
\frac{\gamma }{2};\frac{3}{2};1\right) \right.} \,.  \label{any26.3}
\end{equation}

Finally, taking into account the symmetries of the several homogeneous and
nonhomogeneous expectation values under the changes $x_{1}\leftrightarrow
y_{1}$, $x_{2}\leftrightarrow y_{2}$, $x_{1}\leftrightarrow x_{2}$, $%
y_{1}\leftrightarrow y_{2}$ the full expectation value of the potential (\ref%
{any22.1}) has the form (replacing $\lambda $ by $\Omega /\hbar $),%
\begin{eqnarray}
\left\langle V\right\rangle &=&\frac{16}{3}\left\langle
Q_{x_{1}}^{4}\right\rangle +4\left\langle
Q_{x_{1}}^{2}Q_{x_{2}}^{2}\right\rangle =\frac{\hbar ^{2}\left( \gamma
+3\right) \left( \gamma +2\right) }{6\Omega ^{2}}\left( \left.
_{2}F_{1}\left( \frac{1-\gamma }{2},-\frac{\gamma }{2};\frac{3}{2};1\right)
\right. \right) ^{-1}  \notag \\
&\times &\left\{ 5\left. _{2}F_{1}\left( \frac{1-\gamma }{2},-\frac{\gamma }{%
2};\frac{5}{2};1\right) \right. +\frac{\gamma \left( \gamma -1\right) }{5}%
\left. _{2}F_{1}\left( \frac{3-\gamma }{2},\frac{2-\gamma }{2};\frac{7}{2}%
;1\right) \right. \right\} \,.  \label{any27}
\end{eqnarray}

\section{Selected Hamiltonian operators}

Traditionally the Hamiltonian operator for particles with fractional
statistics is represented by the standard nonrelativistic Hamiltonian for a
charged particle interacting with a magnetic flux tube through the minimum
coupling with a vector potential \cite{Wilczek,Wu84}. The latter potential
may conveniently be removed by a gauge transformation which, in particular,
gives the multivalued character to the corresponding wave function \cite%
{Halperin84,Wu84}\footnote{%
Regarding the gauge transformation there is a frequently used terminology
associated with that \cite{IenLec90,IenLec92}. The corresponding \textit{%
singular }gauge transformation defines the so-called \textit{anyons gauge}
in which the wave function is a multivalued function and the Hamiltonian
does not have the potential vector. The other possibility is the so-called
\textit{CS-gauge} where the wave function obeys the standard statistics but
the Lagrangian contains a \textit{Chern-Simons} term. In this paper we do
not follow the latter approach.}. Once we are working with a multivalued
wave function (more preciselly with multivalued coherent states), the
nonrelativistic momentum operator does have such potential vector. Moreover
we add to the previous standard descriptions \cite{Wu84} a quartic potential
of the form (\ref{any22.1}). For us this potential has a great importance
due to its close relation with rotational invariant models.

The nonrelativistic Hamiltonian operator in consideration has the form%
\begin{eqnarray}
&&\mathcal{H}\left( \mathbf{P},\mathbf{Q}\right) =\sum_{\sigma =1}^{2}\left[
\frac{\mathbf{P}_{\sigma }^{2}}{2m}+\frac{m\varpi ^{2}}{2}\mathbf{Q}_{\sigma
}^{2}\right] +gV\,,\ \ V=\left( \mathbf{Q}_{1}^{2}+\mathbf{Q}_{2}^{2}\right)
^{2}\,,  \label{cl1} \\
&&\mathbf{P}_{\sigma }=\left( P_{x_{\sigma }},P_{y_{\sigma }}\right) \,,\ \
\mathbf{Q}_{\sigma }=\left( Q_{x_{\sigma }},Q_{y_{\sigma }}\right) \,,
\notag
\end{eqnarray}%
where $m$ represent the mass of the particles, $\varpi $ is the harmonic
potential frequency, $g$ is the quartic interaction coupling constant. In
all cases the enhanced classical hamiltonian is identified with the
expectation value of the Hamiltonian operator (\ref{cl1}) with respect to
the coherent states,%
\begin{eqnarray}
H\left( \mathbf{p},\mathbf{q}\right)  &=&\left\langle \mathbf{p},\mathbf{q}%
\right\vert \mathcal{H}\left( \mathbf{P},\mathbf{Q}\right) \left\vert
\mathbf{p},\mathbf{q}\right\rangle   \notag \\
&=&\sum_{\sigma =1}^{2}\left[ \frac{\mathbf{p}_{\sigma }^{2}}{2m}+\frac{%
m\varpi ^{2}}{2}\mathbf{q}_{\sigma }^{2}+\frac{\left\langle \mathbf{P}%
_{\sigma }^{2}\right\rangle }{2m}+\frac{m\varpi ^{2}}{2}\left\langle \mathbf{%
Q}_{\sigma }^{2}\right\rangle\right] +g\left\langle {\bf{p}},{\bf{q}}\vert V\vert {\bf{p}},{\bf{q}}\right \rangle \,,
\label{cl2}
\end{eqnarray}%
where the coherent states are defined, as in (\ref{coherstates}), by%
\begin{equation}
\left\vert \mathbf{p},\mathbf{q}\right\rangle =\prod_{\sigma =1}^{2}U\left(
\mathbf{q}_{\sigma }\right) V\left( \mathbf{p}_{\sigma }\right) \left\vert
\eta \right\rangle \,,\ \ U\left( \mathbf{q}_{\sigma }\right) =e^{-i\mathbf{q%
}_{\sigma }\cdot \mathbf{P}_{\sigma }/\hbar }\,,\ \ V\left( \mathbf{p}%
_{\sigma }\right) =e^{i\mathbf{p}_{\sigma }\cdot \mathbf{Q}_{\sigma }/\hbar
}\,.  \label{cl0}
\end{equation}%
As one can see from (\ref{cl2}) there is no fundamental\footnote{%
More precisely speaking, the functional form of the enhanced Hamiltonian
for bosons, fermions and anyons are the same. The difference appears only in
the numerical values of the expectation values which, as a matter of fact,
depend on the choice of fiducial vectors under consideration.} difference
for the enhanced classical Hamiltonian between bosons, fermions or anyons.
Nevertheless there does exist quantum differences between them, or more
precisely, the numerical values of the coefficients proportional to $\hbar $
have a different value for each considered particle. By virtue of (\ref{kla5}%
) and symmetries behind the several expectation values of (\ref{any22.2}),
as discussed in the subsection above, the expectation value of the potential
$V$ with respect to the coherent states (\ref{cl0}) reads%
\begin{eqnarray}
\left\langle \mathbf{p},\mathbf{q}\right\vert V\left\vert \mathbf{p},\mathbf{%
q}\right\rangle  &=&\sum_{\sigma =1}^{2}\left[ \left( \mathbf{q}_{x_{\sigma
}}^{2}+\mathbf{q}_{y_{\sigma }}^{2}\right) ^{2}+10\,\mathbf{q}_{\sigma
}^{2}\left\langle Q_{x_{1}}^{2}\right\rangle \right]   \notag \\
&+&2\left( \mathbf{q}_{1}\cdot \mathbf{q}_{2}\right) \left( \left\langle
Q_{x_{1}}Q_{x_{2}}\right\rangle +\left\langle
Q_{y_{1}}Q_{y_{2}}\right\rangle \right) +\frac{16}{3}\left\langle
Q_{x_{1}}^{4}\right\rangle +4\left\langle
Q_{x_{1}}^{2}Q_{x_{2}}^{2}\right\rangle \,.  \label{cl2.2}
\end{eqnarray}

Labeling $H_{k}\left( \mathbf{p},\mathbf{q}\right) $ as the enhanced
Hamiltonian for bosons $\left( k=b\right) $, fermions $\left( k=f\right) $, $%
\left( k=\gamma \right) $ for anyons and using the results of the previous
subsections we list below the enhanced hamiltonian
for bosons and fermions:
\begin{eqnarray}
H_{b}\left( \mathbf{p},\mathbf{q}\right)  &=&H_{c}\left( \mathbf{p},\mathbf{q%
}\right) +\hbar \sum_{\sigma =1}^{2}\left( \frac{3g}{\Omega }\right) \mathbf{%
q}_{\sigma }^{2}+\hbar \left( \frac{\Omega }{m}+\frac{m\varpi ^{2}}{\Omega }%
\right) +\hbar ^{2}\left( \frac{3g}{\Omega ^{2}}\right) \,,  \label{cl3} \\
H_{f}\left( \mathbf{p},\mathbf{q}\right)  &=&H_{c}\left( \mathbf{p},\mathbf{q%
}\right) +6\,\hbar \sum_{\sigma =1}^{2}\left( \frac{3g}{\Omega }\right)
\mathbf{q}_{\sigma }^{2}+\hbar \left( \frac{3\Omega }{2m}+\frac{3m\varpi ^{2}%
}{2\Omega }\right) +2\,\hbar ^{2}\left( \frac{3g}{\Omega ^{2}}\right) \,,
\label{cl4}
\end{eqnarray}%
 and for anyons we have%
\begin{eqnarray}
H_{\gamma }\left( \mathbf{p},\mathbf{q}\right)  &=&H_{c}\left( \mathbf{p},%
\mathbf{q}\right) +\hbar \left[ \frac{\Omega }{m}\left( 1+\gamma \frac{%
\left. _{2}F_{1}\left( \frac{2-\gamma }{2},\frac{1-\gamma }{2};\frac{3}{2}%
;1\right) \right. }{\left. _{2}F_{1}\left( \frac{1-\gamma }{2},-\frac{\gamma
}{2};\frac{3}{2};1\right) \right. }-\frac{\gamma }{2}\right) +\frac{m\varpi
^{2}}{2\Omega }\left( 2+\gamma \right) \right]   \notag \\
&+&\frac{\hbar g\left( \gamma +2\right) }{\Omega }\left[ \sum_{\sigma =1}^{2}%
\frac{5\mathbf{q}_{\sigma }^{2}}{2}-\left( \mathbf{q}_{1}\cdot \mathbf{q}%
_{2}\right) \left( \frac{\left. _{2}F_{1}\left( -\frac{1+\gamma }{2},-\frac{%
\gamma }{2};\frac{3}{2};1\right) \right. }{\left. _{2}F_{1}\left( \frac{%
1-\gamma }{2},-\frac{\gamma }{2};\frac{3}{2};1\right) \right. }-1\right) %
\right]   \notag \\
&+&\frac{\hbar ^{2}\left( \gamma +3\right) \left( \gamma +2\right) }{6\Omega
^{2}}\left( \left. _{2}F_{1}\left( \frac{1-\gamma }{2},-\frac{\gamma }{2};%
\frac{3}{2};1\right) \right. \right) ^{-1}  \notag \\
&\times &\left\{ 5\left. _{2}F_{1}\left( \frac{1-\gamma }{2},-\frac{\gamma }{%
2};\frac{5}{2};1\right) \right. +\frac{\gamma \left( \gamma -1\right) }{5}%
\left. _{2}F_{1}\left( \frac{3-\gamma }{2},\frac{2-\gamma }{2};\frac{7}{2}%
;1\right) \right. \right\} \,.  \label{cl5}
\end{eqnarray}%
where the \textit{classical Hamiltonian }$H_{c}\left( \mathbf{p},\mathbf{q}%
\right) $, in which $\hbar \rightarrow 0$, has the same form for all cases,%
\begin{equation}
H_{c}\left( \mathbf{p},\mathbf{q}\right) =\lim_{\hbar \rightarrow
0}H_{k}\left( \mathbf{p},\mathbf{q}\right) =\sum_{\sigma =1}^{2}\left( \frac{%
\mathbf{p}_{\sigma }^{2}}{2m}+\frac{m\varpi ^{2}}{2}\mathbf{q}_{\sigma
}^{2}+g\left( \mathbf{q}_{x_{\sigma }}^{2}+\mathbf{q}_{y_{\sigma
}}^{2}\right) ^{2}\right) \,.  \label{cl7}
\end{equation}

\section{Conclusion}
In this paper we have focussed on a central question in enhanced quantization using canonical coherent states, namely, the choice of the fiducial vector and the issues that choice involves. Initially, it was argued that a good choice is largely dictated by the explicit form of the Hamiltonian operator under consideration, and, in many cases the choice of the unique ground state as the fiducial vector has several virtues. However, that choice can also be relaxed to consider other fiducial vectors, and we can
illustrate that choice by focussing attention on a Hamiltonian operator with a quartic interaction. One reason behind this choice is basically due to the fact that it is, effectively, the simplest example in which ${\cal O}(\hbar)$ coefficients of dynamical terms  are involved that modify the classical description. Secondly, this choice can exhibit a problem with symmetry, e.g., rotational invariance,
 and similar properties can be extended to other models. In particular, fiducial vectors based on  a Gaussian form are appropriate for bosons to deal with these Hamiltonian operators and $\hbar$-dynamical coefficients can be consistently treated with them. Generally, such coefficients can also be reduced by choosing Gaussian fiducial vectors, although this is not required according to the principles of enhanced canonical quantization \cite{EQ, Kla63I, Kla63II, Kla63III}.
In such cases, the corresponding enhanced Hamiltonian is a symbol \cite{Husimi} of the respective Hamiltonian operator where $\hbar$-dependence is included.

 Although commonly used, it is a fact that Gaussian fiducial vectors are not always suitable to consistently describe certain physical systems. In this respect, we have chosen two examples where this form clearly fails: the enhanced quantization of fermions and anyons. In these cases the fiducial vectors cannot be independent Gaussians, and, instead, they must involve cross correlations for fermions or snyons. This latter property has a necessary physical consequence, namely, ensuring that permutations of the variables  of the coherent state representation of Hilbert space vectors involve the required change of phase.

In section 3 non-Gaussian fiducial vectors have been used in the coherent states constructed for fermions and anyons. We have calculated several expectation values for both systems and, in this regard, the exact calculations for anyons also include the corresponding results for fermions simply by choosing $\gamma =1$. Using these results, we have calculated the enhanced classical Hamiltonian and after taking the limit in which $\hbar\ra0$, we have shown that bosons, fermions, and anyons all have the same classical Hamiltonian,  despite the fundamental differences between their properties when $\hbar>0$.

\section*{Appendix}

In dealing with expectation values\ of coordinate and momentum operators for
the anyon case we frequently encounter integrals of the form [recall the convention: $\tint_a^b\s(\cdot)\s d{\bf r}=\tint_a^b(\cdot)\s dx\,\tint_a^b(\cdot)\s dy$]
\begin{equation}
\int_{-\infty }^{\infty }d\mathbf{r}_{\sigma ^{\prime }}\left\vert \mathbf{r}%
_{\sigma ^{\prime }}-\mathbf{r}_{\sigma }\right\vert ^{2\gamma }e^{-\lambda
\mathbf{r}_{\sigma ^{\prime }}^{2}}\,,  \label{ap1}
\end{equation}%
where the indexes $\sigma ,\sigma ^{\prime }$ label particles. Due to
rotational symmetry, one can choose a particular reference system to
simplify its solution. Choosing, for example, a reference frame where the
vector $\mathbf{r}_{\sigma }$ is aligned along the axis $y_{\sigma ^{\prime
}}$ we have $\mathbf{r}_{\sigma }\cdot \mathbf{r}_{\sigma ^{\prime
}}=r_{\sigma }r_{\sigma ^{\prime }}\cos \left( \pi /2-\vartheta \right)
=r_{\sigma }r_{\sigma ^{\prime }}\sin \vartheta $. Here $\vartheta $ is the
angle between the axis $x_{\sigma ^{\prime }}$ and $\mathbf{r}_{\sigma
^{\prime }}$ such that (\ref{ap1}) admits the form \cite{Gradshtein},%
\begin{eqnarray}
\rho _{\gamma }^{\left( \pm \right) }\left( r_{\sigma }\right)
&=&\int_{0}^{\infty }dr_{\sigma ^{\prime }}r_{\sigma ^{\prime }}^{2\pm
1}e^{-\lambda r_{\sigma ^{\prime }}^{2}}\mathcal{I}_{\vartheta }^{\left( \pm
\right) }\left( r_{\sigma },r_{\sigma ^{\prime }};\gamma \right) \,,  \notag
\\
\mathcal{I}_{\vartheta }^{\left( \pm \right) }\left( r_{\sigma },r_{\sigma
^{\prime }};\gamma \right)  &=&\int_{0}^{2\pi }d\vartheta \left( \cos
\vartheta \right) ^{1\pm 1}\left( r_{\sigma }^{2}+r_{\sigma ^{\prime
}}^{2}-2r_{\sigma }r_{\sigma ^{\prime }}\sin \vartheta \right) ^{\gamma }
\notag \\
&=&\left( \frac{3\mp 1}{2}\right) \pi \left( r_{\sigma }^{2}+r_{\sigma
^{\prime }}^{2}\right) ^{\gamma }\left. _{2}F_{1}\left( \frac{1-\gamma }{2},-%
\frac{\gamma }{2};\frac{3\pm 1}{2};\frac{4r_{\sigma }^{2}r_{\sigma ^{\prime
}}^{2}}{\left( r_{\sigma }^{2}+r_{\sigma ^{\prime }}^{2}\right) ^{2}}\right)
\right. \,.  \label{ap2}
\end{eqnarray}%
To evaluate $\rho _{\gamma }^{\left( \pm \right) }\left( r_{\sigma }\right) $
it is convenient to make use the series representation of the hypergeometric
functions \cite{Gradshtein},%
\begin{equation}
\left. _{2}F_{1}\left( \frac{1-\gamma }{2},-\frac{\gamma }{2};\frac{3\pm 1}{2%
};\frac{4r_{\sigma }^{2}r_{\sigma ^{\prime }}^{2}}{\left( r_{\sigma
}^{2}+r_{\sigma ^{\prime }}^{2}\right) ^{2}}\right) \right.
=\sum_{s=0}^{\infty }\frac{1}{s!}\left( \frac{1-\gamma }{2}\right)
_{s}\left( -\frac{\gamma }{2}\right) _{s}\left( \frac{3\pm 1}{2}\right)
_{s}^{-1}\frac{\left( 4r_{\sigma }r_{\sigma ^{\prime }}\right) ^{2s}}{\left(
r_{\sigma }^{2}+r_{\sigma ^{\prime }}^{2}\right) ^{2s}}\,,  \label{ap2.1}
\end{equation}%
which, after the change of variables $u_{\sigma }=\lambda r_{\sigma }^{2}$
and $u_{\sigma ^{\prime }}=\lambda r_{\sigma ^{\prime }}^{2}$, can be
brought into the form%
\begin{eqnarray}
&&\rho _{\gamma }^{\left( \pm \right) }\left( u_{\sigma }\right)
=\sum_{s=0}^{\infty }\Lambda _{s}^{\left( \pm \right) }\varrho _{\gamma
,s}^{\left( \pm \right) }\left( u_{\sigma }\right) u_{\sigma }^{s}\,,  \notag
\\
&&\varrho _{\gamma ,s}^{\left( \pm \right) }\left( u_{\sigma }\right) \equiv
\int_{0}^{\infty }du_{\sigma ^{\prime }}e^{-u_{\sigma ^{\prime }}}u_{\sigma
^{\prime }}^{s+\frac{1\pm 1}{2}}\left( u_{\sigma ^{\prime }}+u_{\sigma
}\right) ^{\gamma -2s}\,,  \notag \\
&&\Lambda _{s}^{\left( \pm \right) }=\frac{\pi }{\lambda ^{\gamma +\frac{%
3\pm 1}{2}}}\frac{4^{s}}{s!}\left( \frac{1-\gamma }{2}\right) _{s}\left( -%
\frac{\gamma }{2}\right) _{s}\left( \frac{3\pm 1}{2}\right) _{s}^{-1}\,,
\label{ap3}
\end{eqnarray}%
where $\left( n\right) _{s}=\Gamma \left( n+s\right) /\Gamma \left( s\right)
$ is the Pochammer symbol \cite{Gradshtein}. The integral above can be
analytically solved,%
\begin{eqnarray}
&&\varrho _{\gamma ,s}^{\left( \pm \right) }\left( u_{\sigma }\right)
=\Gamma \left( s+\frac{3\pm 1}{2}\right) u_{\sigma }^{\frac{2\left( \gamma
-s\right) +1\pm 1}{4}}e^{\frac{u_{\sigma }}{2}}W_{\mu ,\nu }\left( u_{\sigma
}\right) \,,  \notag \\
&&\mu =\frac{2\left( \gamma -3s\right) -1\mp 1}{4}\,,\ \ \nu =\frac{2\left(
s-\gamma \right) -3\mp 1}{4}\,,  \label{ap4}
\end{eqnarray}%
with $W_{\mu ,\nu }\left( u_{\sigma }\right) $ being a Wittaker function
\cite{Gradshtein}.

To complete the description it is also needed to solve one more integral of
the expressions above, namely an integral over the variables $u_{\sigma }$
\cite{Gradshtein},%
\begin{align}
R_{n}^{\left( \pm \right) }\left( \gamma ,s\right) &\equiv \int_{0}^{\infty
}du_{\sigma }e^{-\frac{u_{\sigma }}{2}}u_{\sigma }^{\frac{\gamma +s}{2}%
+n}W_{\mu ,\nu }\left( u_{\sigma }\right)\notag \\
&=\frac{\Gamma \left( n+s+\frac{%
3\mp 1}{4}\right) \Gamma \left( n+\gamma +\frac{9\pm 1}{4}\right) }{\Gamma
\left( 2s+n+\frac{9\pm 1}{4}\right) }\,,\ \ n>\frac{-3\pm 1}{4}\,.
\label{ap5.1}
\end{align}

\section*{Acknowledgements}

T. C. Adorno acknowledges support of FAPESP under the contracts 2013/00840-9
and 2013/16592-4. He is also thankful to the Dept. of Physics of the
University of Florida for its kind hospitality.

\end{document}